\documentclass{emulateapj}

\newcommand{\kms}{km~s$^{-1}$}

\newcommand{\ldl}{$\lambda/{\Delta}{\lambda}$}
\newcommand{\teff}{T$_{eff}$}

\newcommand{\water}{H$_2$O}
\newcommand{\name}{DENIS~J220002.05$-$303832.9}
\newcommand{\namesh}{DENIS~2200$-$3038}

\slugcomment{Accepted to AJ}

\shorttitle{DENIS~2200$-$3038AB}
\shortauthors{Burgasser \& McElwain}

\begin{document}

\title{Resolved Spectroscopy of M Dwarf/L Dwarf Binaries. I. DENIS~J220002.05$-$303832.9AB}

\author{Adam J.\ Burgasser\altaffilmark{1}}
\affil{Massachusetts Institute of Technology, Kavli Institute for Astrophysics and Space Research,
77 Massachusetts Avenue, Building 37,
Cambridge, MA 02139-4307, USA; ajb@mit.edu}
\and
\author{Michael W.\ McElwain}
\affil{University of California Los Angeles, Division of Astronomy \& Astrophysics, 8965
Math Science Bldg., 405 Hilgard Ave., Los Angeles, CA, 90095-1562;
mcelwain@astro.ucla.edu}

\altaffiltext{1}{Visiting Astronomer at the Infrared Telescope Facility, which is operated by
the University of Hawaii under Cooperative Agreement NCC 5-538 with the National Aeronautics
and Space Administration, Office of Space Science, Planetary Astronomy Program.}

\begin{abstract}
We present the discovery of the common proper motion
M9 + L0 binary DENIS~J220002.05$-$303832.9AB, identified
serendipitously with the SpeX near infrared imager/spectrograph.
Spectral types are derived from resolved near infrared spectroscopy of the
well-separated (1$\farcs$09$\pm$0$\farcs$06) components and comparison to
equivalent data for M and L dwarf spectral standards.
Physical association is deduced from the angular proximity of the sources,
their common proper motion
and their similar spectrophotometric distances (35$\pm$2~pc).
The estimated distance of this pair implies a projected separation of
38$\pm$3~AU, wider than typical separations
for other M dwarf/L dwarf binaries, but consistent with the
maximum separation/total system mass trend previously identified by Burgasser et al.\ (2003).
We discuss the DENIS~2200$-$3038AB system in context with other low mass binaries,
and its role in
studying dust formation processes and activity trends across
the transition between the M and L dwarf spectral classes.
\end{abstract}


\section{Introduction}

Multiple stellar systems are important probes of star formation processes and
atmospheric physics, and provide one of the few means of directly measuring
stellar mass.  Multiples are particularly useful for studying the physical properties
and origins of the lowest mass stars and brown dwarfs, the latter being
stars of such low mass that they are incapable of sustaining core hydrogen
fusion \citep{kum62,hay63}.
Coeval binaries facilitate the study of surface gravity and temperature
effects on the complex spectra of cool dwarfs independent of age or metallicity variations,
and mass measurements allow specific tests of evolutionary theory \citep{lan01,bou04,zap04}.
The multiplicity fraction, separation distribution and mass ratio distribution of
low mass binaries provide critical constraints on the current menagerie of
formation models \citep{bou03,me03b,clo03}.

High resolution imaging surveys of very low mass
stellar and brown dwarf systems (M$_{total} \lesssim$ 0.2 M$_{\sun}$)
have been conducted by several groups in recent years, focusing on both
field \citep{koe99,mar99,rei01,clo02,clo03,bou03,me03b,giz03,sie03,sie05}
and young cluster \citep{mar98,mar00b,mar03,neu02,pin03,ken05,kra05,luh05b} populations.
These studies have converged to similar results: very low mass binaries
are relatively rare ($f_{bin} \sim$ 10--20\%) and
tend to form closely-separated ($\rho$ $\lesssim$ 20~AU), nearly-equal
mass ($q \equiv {\rm M}_B/{\rm M}_A \sim$ 1) systems.
These results lend support for dynamical
ejection formation models
\citep{ste98,ste03,rpt01,bat02a,bat02b,bat03,del03,del04,bat05,umb05},
in which gravitational scattering of protostellar cores prevents
both the accretion of significant gas and dust (resulting in low masses)
and the integrity of any weakly bound low-mass systems.

However, a handful of widely separated low mass binaries
have been identified serendipitously, including
GG~Tau~Bab \citep{lei91,whi99},
2MASS~J11011926-7732383AB \citep[hereafter 2MASS~1101-7732AB]{luh04}
and DENIS~J055146.0-443412.2AB \citep[hereafter DENIS~0551-4434AB]{bil05}
These systems, containing late-type M and (in the case of DENIS~0551$-$4434AB)
L dwarf components, have projected separations of over 200~AU,
more than 10 times wider than the apparent separation
limit found in high resolution imaging surveys.
The first two binaries, identified in the young associations
Taurus Auriga and Chameleon I, contain substellar
mass components.  DENIS~0551-4434AB is likely composed of two low mass stars.
These systems are fragile, with escape velocities $V_{esc} \lesssim 1$~{\kms},
smaller than the $\sim$2~{\kms} velocity dispersions observed in young clusters
\citep{jon88,joe01} and predicted in ejection models \citep{bat03}.
Both \citet{luh04} and \citet{bil05} argue that the existence of such
weakly bound systems provides strong evidence that dynamic processes
are not the exclusive means of forming the lowest mass stars and brown dwarfs.

In this article, we present the discovery of a new low mass binary,
{\name}AB (hereafter {\namesh}).
Originally identified as an unresolved
L0 dwarf by \citet{ken04} in the Deep Near Infrared Survey of the
Southern Sky \citep{epc97}, we find this source to be a well-separated (1$\farcs$1, $\sim$40~AU) M9 + L0 pair.
In $\S$~2 we describe our imaging and spectroscopic observations, and
data reduction techniques.
In $\S$~3 we present analysis of the data, including resolved photometry
and spectroscopy of the pair and its systemic properties.
Physical association is deduced from the angular proximity,
similar spectrophotometric distances and common proper motion
of the components.
In $\S$~4 we discuss the properties of this system in context with other low mass binaries
and its role in the study of atmospheric and activity trends
across the M dwarf/L dwarf boundary.
Results are summarized in $\S$ 5.

\section{Observations}

\subsection{Imaging}

{\namesh} was observed on 2004 September 7 (UT)
as part of a campaign to acquire low resolution
near infrared spectra of late-type comparison stars using the
SpeX spectrograph \citep{ray03}, mounted on the
3m NASA Infrared Telescope Facility (IRTF).
Conditions during the observations were non-photometric,
with light cirrus but excellent seeing
(0$\farcs$6 at $J$-band).
Acquisition images with the instrument's guiding camera revealed
two distinct sources at the position of {\namesh},
separated by roughly 1$\arcsec$.  We
obtained a series of images of the pair through the Mauna Kea Observatory (MKO)
$JHK$ filter set \citep{sim02,tok02},
interspersed with images of a nearby
bright single star 2MASS~J22001305-3041415
(2MASS $J$ = 12.51$\pm$0.03, $J-K_s$ = 0.41$\pm$0.04) as a
point spread function (PSF) calibration source.
The image rotator was set to 0$\degr$.
Eight dithered exposures were obtained for each pointing,
for a total of 200, 120 and 120~s integration for {\namesh}
at $J$, $H$ and $K$, respectively.
We also obtained twilight and bias exposures two nights later
during the same observing campaign.

\begin{figure}
\epsscale{1.0}
\plotone{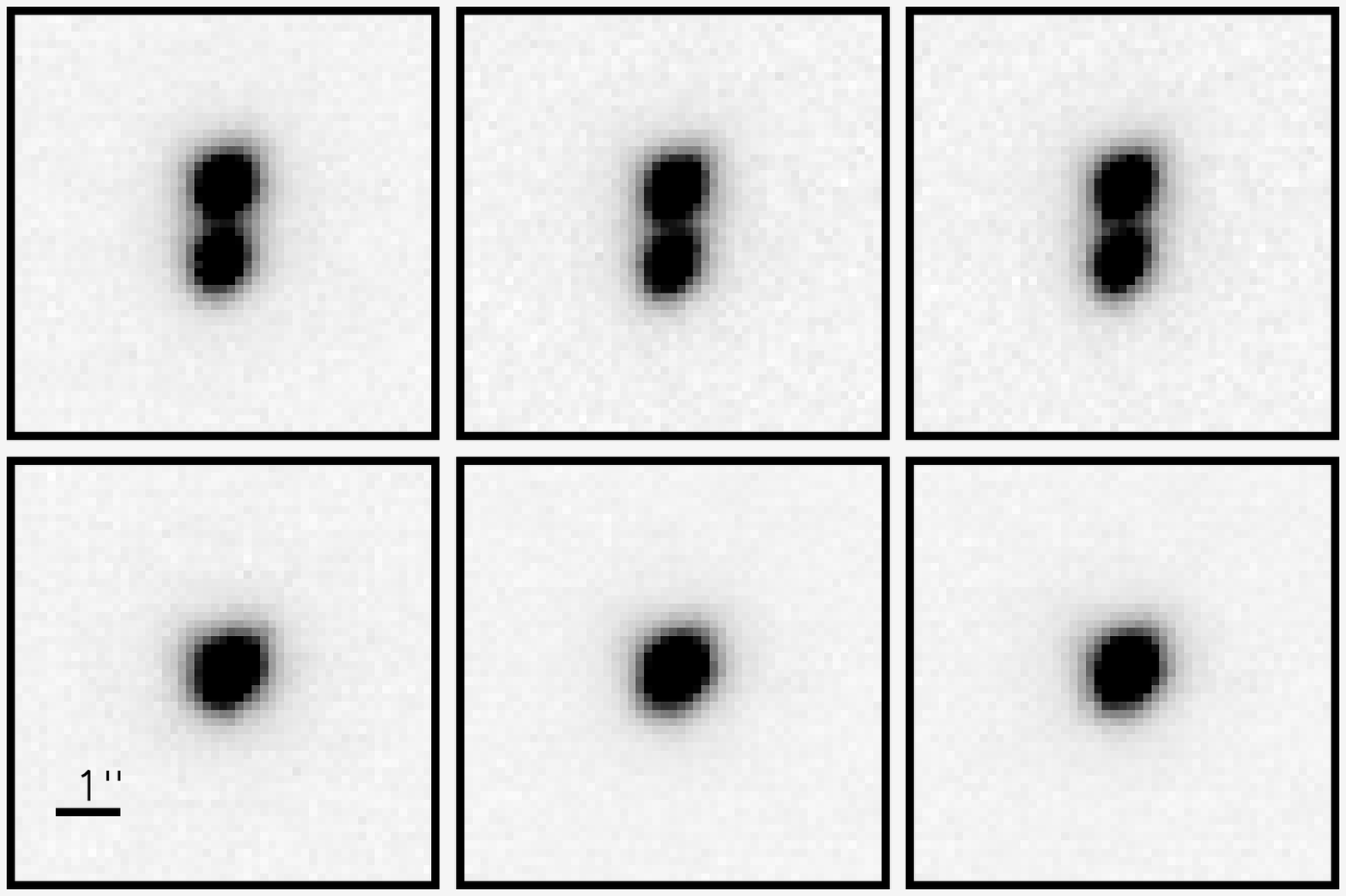}
\caption{Reduced mosaic images of {\name}AB (top) and the PSF calibrator
2MASS~J22001305$-$3041415 (bottom) in the $J$- (left), $H$- (middle) and $K$-bands (right).
Images are 6$\arcsec$ on a side, oriented with north at the top and east to the left.
\label{fig1}}
\end{figure}

The imaging data were cleaned using a pixel
mask constructed from the bias and twilight frames,
then pair-wise subtracted and divided by the appropriate flat field image
for pixel response calibration.  The flat field images
were created by bias-subtracting, median-combining and normalizing the twilight exposures.
The calibrated images were centered and coadded
to produce a mosaic for each filter/pointing combination.
Figure~\ref{fig1} displays $6{\arcsec}{\times}6{\arcsec}$
subsections of the $JHK$ mosaics for the target and PSF star.
The two {\namesh} sources are clearly resolved, lying along a north-south
axis. The northern source is slightly brighter in all three bands.

We measured the relative astrometry and photometry of the {\namesh} pair
using an iterative PSF fitting algorithm, as follows.
First, for each filter the peak positions
and fluxes of the two sources
were estimated by summing the columns and rows
in the appropriate mosaic image.  A model image was then constructed from two
of the reduced PSF images acquired in the same filter, shifting and scaling these images to match the
estimated positions and fluxes of the {\namesh} sources.  This model image was then subtracted from
one of the reduced {\namesh} images, and residuals (standard deviation in the image) computed.
The primary pixel coordinate, secondary pixel
coordinate, primary flux and secondary flux in the model image were then adjusted iteratively
in steps of 0.1 pixels and 0.01 fraction flux
until residuals were minimized, typically of order 1--2\% of the source peak flux.
This procedure was repeated
for each reduced image in a target/PSF pointing set, yielding 32 fits per filter.
The mean of the flux ratios\footnote{Note that the imaging observations
were taken in non-photometric conditions; however, we assume that the
relative atmospheric transmission over the small separation of the {\namesh} pair
was constant, and that atmospheric absorption has minimal effect
on the relative flux within each individual filter.}, separations ($\rho$) and position angles ($\phi$) for each filter
were adopted as the derived values, and their scatter as an estimate of experimental uncertainty.
We also included uncertainty in the pixel scale (0$\farcs$120$\pm$0$\farcs$002 pixel$^{-1}$)
and rotator alignment (0$\fdg$25) in the error budget (J.\ Rayner, 2005, private communication).
Results are listed in Table~\ref{tab1}.

To calibrate our fitting routine, we repeated the analysis
using simulated binary images constructed
from random pairings of the PSF images.  The simulated images were constructed so as to match
the derived binary parameters for {\namesh} for each filter, with some variation
to replicate the experimental uncertainties.
These images were run through the same routine using a third randomly selected PSF image
for the fit, and the derived parameters compared to the input values.
A total of 10~000 trials were performed for each filter.  Overall, no significant
systematic deviations were found in these simulations; however, scatter in the simulated fits
was significantly larger than the scatter derived from fits to the data.
We adopt the standard deviation among the simulation values as estimates
of the systematic uncertainty in our fitting technique; these are listed separately in
Table~\ref{tab1}.

Second epoch $J$-band images of the {\namesh} pair and
the PSF star were also obtained with SpeX exactly one year later on 2005 September 7 (UT).
Conditions during this run were poor, with seeing of 0$\farcs$9.
Six 10~s dithered exposures were obtained of both fields, and analyzed using
the same fitting routine.  These observations are discussed in further detail in $\S$~3.2.

\subsection{Spectroscopy}

Spectral data for the two {\namesh} sources were
obtained on 2004 September 7 (UT) using the SpeX prism dispersed mode, which
provides low resolution 0.7--2.5~$\micron$ spectra in a single order.
For all observations, the 0$\farcs$5 slit was employed, yielding a
spectral resolution {\ldl} $\approx 150$
and dispersion across the chip of 20--30~{\AA}~pixel$^{-1}$.
Spectra were acquired by setting the image rotator to 90$\degr$, so that
one source lay in the slit while the other source was positioned
orthogonal to the orientation of the slit and used for guiding.
Note that the slit was not aligned with the parallactic angle ($\sim$15$\degr$),
so some differential color refraction is expected in the reduced spectrum (see below).
Six exposures of 120~s each were obtained for both sources
in an ABBA dither pattern along the slit, for a total of 720~s integration
per source.
The A0~V HD~202025 was observed immediately
afterward, at a similar airmass (1.64) and with the slit aligned to the parallactic angle,
for flux calibration.
Internal flat field and Ar arc lamps were observed
for pixel response and wavelength calibration.

All spectral data were reduced using the SpeXtool package, version 3.2
\citep{cus04} using standard settings.
First, the images were corrected for linearity, pair-wise subtracted, and divided by the
corresponding median-combined flat field image.  Spectra were optimally extracted using the
default settings for aperture and background source regions, and wavelength calibration
was determined from arc lamp and sky emission lines.  Multiple spectral observations for each
source were then median-combined after scaling the individual
spectra to match the highest signal-to-noise
observation.  Telluric and instrumental response corrections for the science data were determined
using the method outlined
in \citet{vac03}.  Line shape kernels were derived from the
arc lines.  Adjustments were made to the telluric spectra to compensate
for differing \ion{H}{1} line strengths in the observed A0~V spectrum
and pseudo-velocity shifts\footnote{These shifts, of
order 200~km~s$^{-1}$, compensate for a slight wavelength shift present in the
low resolution Vega model spectrum employed for this mode (M.\ Cushing, 2005, private communcation).}.
Final calibration was made by
multiplying the observed target spectrum by the telluric correction spectrum,
which includes instrumental response correction through the ratio of the observed A0~V spectrum
to a scaled, shifted and deconvolved Kurucz\footnote{\url{http://kurucz.harvard.edu/stars.html}.}
model spectrum of Vega.

Due to the proximity of the two {\namesh} sources, some contamination
of light from one component could be present in the spectrum of the
other component.  To estimate this, we calculated the integrated light
contribution across the 0$\farcs$5 slit for a source separated from
the slit center by 1$\farcs$1, assuming a wavelength-independent\footnote{Seeing
measurements on 2004 September 7 (UT) for the PSF star in the $JHK$ bands were all
0$\farcs$6, so this assumption is valid.}
Gaussian PSF with a seeing full width at half maximum of 0$\farcs$6.
The total contribution is 0.4\% of the peak source flux, implying
contamination of the northern (southern) spectrum by the
southern (northern) source of only 0.3\% (0.5\%) based on the observed flux ratios.
We consider this level of contamination to be negligible for our analysis.

\begin{figure}
\epsscale{0.9}
\plotone{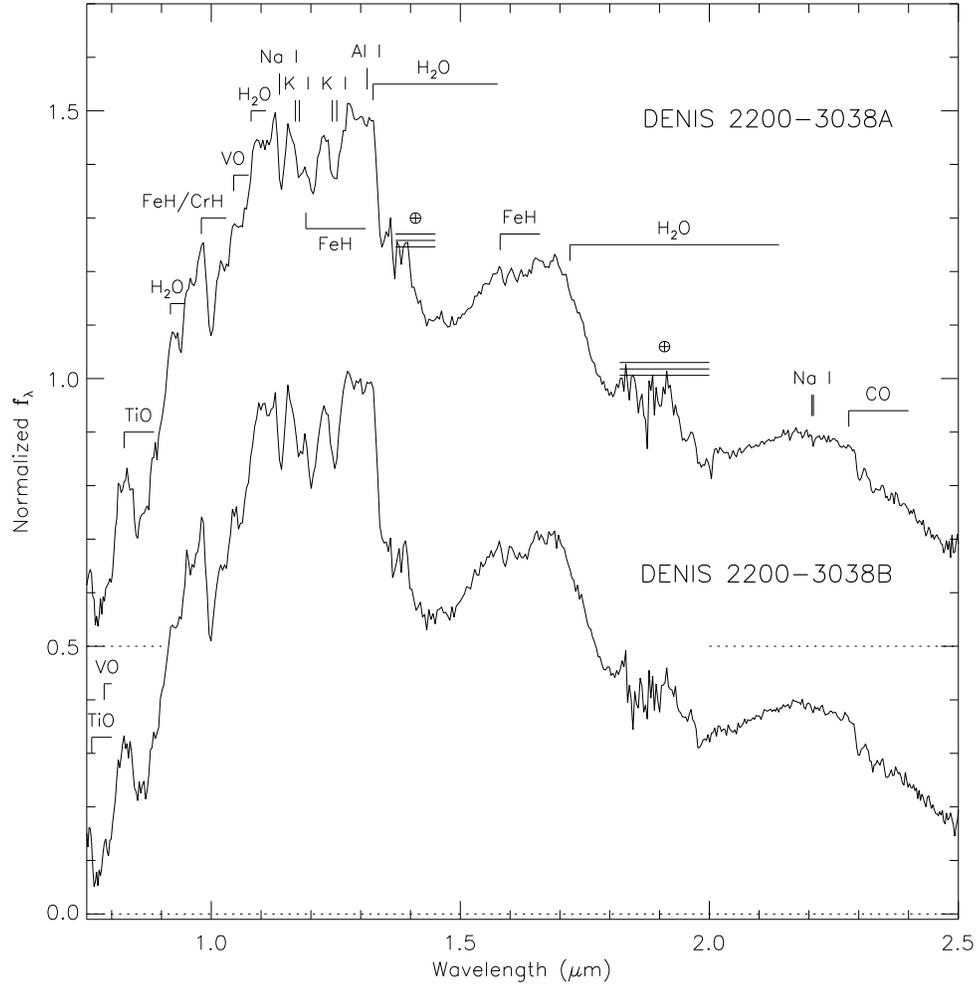}
\caption{Near infrared SpeX spectra of {\namesh}A (top) and B (bottom).
Data are normalized at 1.28~$\micron$, with {\namesh}A offset
by a constant for clarity.  Major molecular (TiO, VO, FeH, CrH, {\water}, CO) and atomic
(\ion{Na}{1}, \ion{K}{1}, \ion{Al}{1}) absorption features are labelled,
as well as regions of strong telluric absorption ($\oplus$).
\label{fig2}}
\end{figure}

The reduced spectra of the two {\namesh} components
are shown in Figure~\ref{fig2}.  Both exhibit near infrared spectral energy
distributions consistent with late-type M and early-type L dwarfs,
including red optical/near infrared spectral slopes;
deep {\water} absorption bands at 1.4 and 1.9~$\micron$;
strong CO absorption at 2.3~$\micron$; FeH bands
at 0.9, 1.2 and 1.6~$\micron$;
TiO and VO absorption at 0.86, 0.78 and 1.05~$\micron$;
and numerous unresolved atomic lines in the $J$- and $K$-bands.
Further discussion on the near infrared spectral characteristics of
late-type M and L dwarfs
can be found in
\citet{jon94};
\citet{leg00,leg01};
\citet{rei01}; \citet{tes01};
\citet{geb02}; \citet{gor03}; \citet{mcl03};
\citet{kna04};
\citet{nak04} and \citet{cus05}.

\section{Analysis}

\subsection{Spectral Classification}

\begin{figure}
\epsscale{1.0}
\plotone{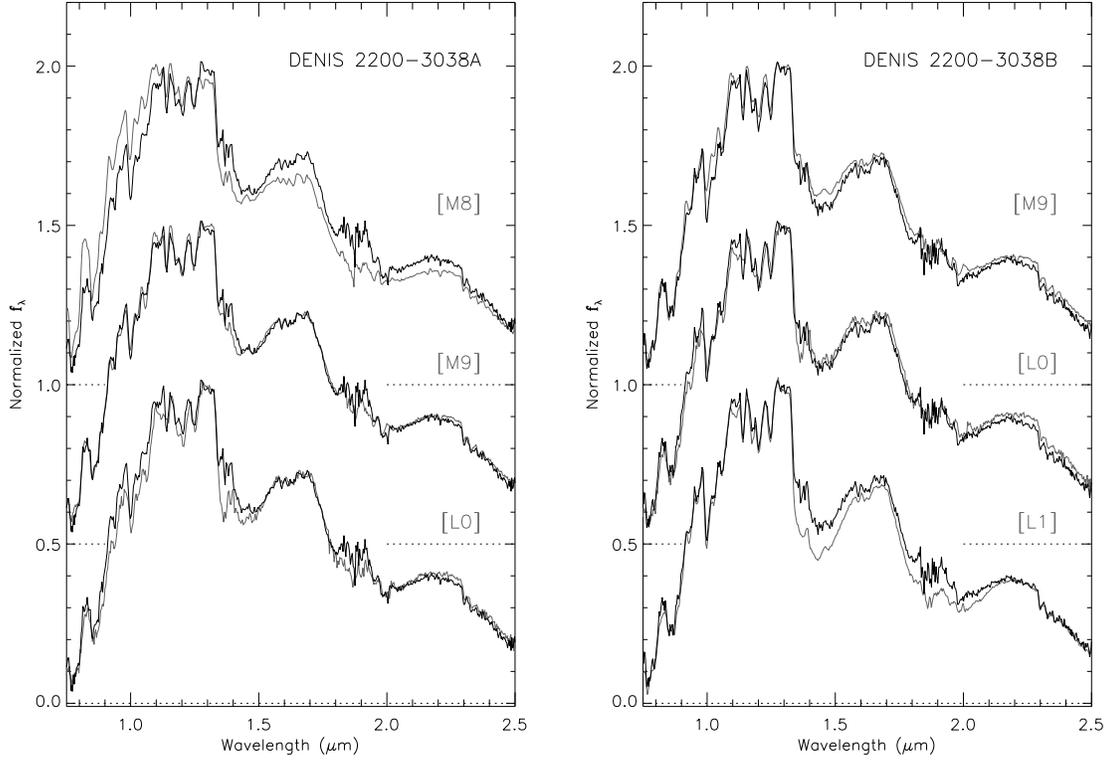}
\caption{Comparison of the near infrared spectra of {\namesh}A and B
(heavy black lines in left and right
panels, respectively)
to equivalent data (grey lines) for the optical spectral standards VB~10 (M8), LHS~2924 (M9),
2MASS~J03454316+2540233 (L0) and 2MASS~J14392836+1929149 (L1).  All spectra have been normalized
at 1.28~$\micron$ and offset by constants.
\label{fig3}}
\end{figure}

Ideally, the classification of a stellar spectrum is most accurately accomplished
by direct comparison of that spectrum to a sequence of pre-defined spectral standards,
with data obtained over a similar waveband and resolution \citep{cor94}.
Currently, no sequence of {\em near infrared} spectral standards exist
for M and L dwarfs, although several studies
have examined methods of classification at these wavelengths
\citep{tok99,rei01,tes01,me02a,geb02,mcl03}.
As a proxy, we compared the {\namesh} spectra to equivalent SpeX prism data
(Burgasser et al.\ 2004b; Burgasser et al.\ in prep.; Cruz et al.\ in prep.)
for the optical spectral standards \citep{kir91,kir99}
VB~10 (M8), LHS~2924 (M9), 2MASS~J03454316+2540233 (L0; hereafter 2MASS~0345+2540)
and 2MASS~J14392836+1929149 (L1).  Figure~\ref{fig3} shows a normalized spectral sequence of these
standards with the spectra of the {\namesh} sources overlain. The brighter northern source
matches the spectrum of LHS~2924 quite well across the full spectral
range, and in particular shows excellent agreement in the band strengths and
overall spectral shape.
The spectrum of the southern component, on the other hand, has deeper {\water}
and FeH bands and weaker TiO and VO absorption,
and shows more consistency with the spectrum of 2MASS~0345+2540.
There are slight discrepancies in the peak $H$- and $K$-band flux peaks,
however, with the southern {\namesh} source exhibiting a slightly
bluer near infrared spectral slope.  We attribute this to light loss at
longer wavelengths due to differential color refraction.  Indeed,
$J-K_s$ colors synthesized from the spectral data of the
northern and southern components
are 0.07 and 0.12 mags bluer than
those derived from the photometry (see below).

Another common method of classifying cool dwarf spectra is
through the use of spectral indices.  For the {\namesh} sources,
we examined indices defined by \citet{rei01}
and \citet{tes01}, both based on low resolution near infrared spectral data.
The Reid et al.\ study defines several indices measuring the strengths
of {\water} and CO bands and color ratios in late-type M and L dwarf spectra,
and provides spectral type
calibrations anchored to optical classifications
for their {\water}-A and {\water}-B indices
(sampling the 1.1 and 1.4~$\micron$ {\water} bands) and the $K1$ index
of \citet{tok99} over spectral types M8 to L6.
\citet{tes01} define six indices sampling {\water}
bands and color ratios, and provide calibrations for these indices over
types L0 to L8.   Due to apparent color refraction effects
in our spectral data, we considered only the four {\water} indices,
and extrapolated the Testi et al.\ linear spectral type calibrations to
incorporate late M dwarf types.
Table~\ref{tab2} lists the derived values and associated
spectral types for all nine indices.  The Reid et al.\
indices yield mean types of M9 and L0 for the northern and southern
components, respectively;
while the Testi et al. indices yield types of M9.5 and L0
with somewhat larger scatter.
Taking these values in conjunction with the direct spectral comparisons
shown in Figure~\ref{fig3}, we adopt spectral types of M9 and L0 for the
two {\namesh} sources, with an uncertainty of $\pm$0.5 subclasses.

\subsection{Is {\namesh}AB a Bound Pair?}

Since the two sources at the position of {\namesh} have similar brightnesses and
spectral types, it appears quite likely that the pair form a gravitationally
bound system.  To assess this, we examined a number of tests for association.
First, the probability of two similarly-typed low mass stars positioned within $\sim$1--2$\arcsec$
of each other is extremely low.  \citet{cru03} identified 53 M9--L0 dwarfs in their search of
the Two Micron All Sky Survey \citep[hereafter 2MASS]{cut03} spanning 16~350~deg$^{2}$,
implying a surface density of
approximately $3{\times}10^{-3}$~deg$^{-2}$ to a depth of $J \approx 14.5$,
similar to the individual apparent $J$ magnitudes of the two {\namesh} sources
(Table~\ref{tab3}; see below).
The probability that these two sources are aligned by chance is only
$10^{-9}$, and can be ruled out statistically with very high confidence.

We also examined the relative spectrophotometric distances of these two sources
(assuming that they are each single) to assess whether they lie at the
same distance.
Table~\ref{tab2} lists individual apparent magnitudes for each source
on the 2MASS photometric system,
derived from 2MASS $JHK_s$ photometry of the unresolved pair and our measured flux
ratios.  The $J$ magnitudes include a small color correction,
${\Delta}J(2MASS)-{\Delta}J(MKO) = -0.02$, computed by integrating
the 2MASS and MKO filter bandpasses over the observed spectra; color corrections at
$H$ and $K/K_s$ were found to be negligible.
(see also Stephens \& Leggett 2004).
\citet{dah02} find a linear relation between absolute $J$-band
magnitude and spectral type for subtypes M6.5 to L8 of $M_J = 8.38 + 0.341{\times}$SpT,
where SpT(M7) = 7, SpT(L0) = 10, etc.  This implies that two sources lying at the
same distance and differing by a single spectral subclass have
${\Delta}M_J = {\Delta}J = 0.34$.  Our value of ${\Delta}J = 0.33{\pm}0.10$
is therefore consistent with the two sources lying at the same distance
to within 0.5$\pm$5\%.
Spectrophotometric distances from the
absolute magnitude/spectral type relations of \citet{cru03} and \citet{vrb04}
yield consistent results,
and we adopt the mean and standard deviation of all of these values, 35$\pm$2~pc,
as our estimated distance to the {\namesh} pair\footnote{Spectrophotometric
distances derived from the absolute magnitude/spectral
type relations of \citet{tin03} are not in agreement with these values, yielding
$\sim$21 and $\sim$31~pc for the northern and southern {\namesh} sources, respectively. This discrepancy
is due to the fact that the high-order polynomial fits
presented in the Tinney et al.\ study are derived for types L0 and later,
and are poorly constrained for late-type M dwarfs.}.

Finally, we assessed whether the pair share common proper motion
by comparing the relative astrometry between our first
and second epoch images.  The 2005 observations yield $\rho = 1{\farcs}08{\pm}0{\farcs}02$
and $\phi = 177{\pm}2{\degr}$ (not including systematic uncertainties),
consistent with the 2004 observations.  The SuperCosmos Sky Survey
\citep{ham01a,ham01b,ham01c} gives a proper motion for {\namesh}
in Right Ascension and declination of
$\mu_{\alpha}\cos{\delta}$ = $0{\farcs}206{\pm}0{\farcs}016$~yr$^{-1}$ and
$\mu_{\delta}$ = $-0{\farcs}063{\pm}0{\farcs}015$~yr$^{-1}$, respectively.
The observed change in the southern component position
relative to the northern component between the two epochs,
including systematic uncertainties, is
${\Delta}\alpha\cos{\delta}$ = $-0{\farcs}01{\pm}0{\farcs}09$ and
${\Delta}\delta$ = $0{\farcs}01{\pm}0{\farcs}09$, implying common
proper motion at the 98.9\% confidence level (2.3$\sigma$).  If systematic uncertainties
are ignored, common proper motion is secure at the 5$\sigma$ level.

Based on these considerations, we confidently claim that the two {\namesh}
components compromise a physically bound, common proper motion system, which
we refer to hereafter as {\namesh}AB.

\subsection{Physical Properties of the Components}

Using the spectrophotometric distance estimate derived above, we deduce a
projected separation for the {\namesh}AB system of $\rho = 38{\pm}3$~AU.
This is rather wide for a very low mass binary, and implies a long
orbital period.
Component masses were estimated using the solar metallicity evolutionary
models of \citet{bur97} for ages of 0.5, 1 and 5~Gyr;
and adopting effective temperatures of 2400 and 2300~K for the A and B components,
respectively, based on the {\teff}/spectral type relation
of \citet[uncertainties of $\sim$100~K]{gol04}.  Derived mass
estimates range over 0.072--0.085~M$_{\sun}$
and 0.069--0.083~M$_{\sun}$ for the A and B components, respectively,
implying a mass ratio of 0.96--0.98 and a
total system mass of 0.14--0.17~M$_{\sun}$.
Hence, the period of this system, assuming a typical semi-major axis $a$ = 1.29$\rho$
\citep{fis92}, is roughly 750--1000~yr.  {\namesh}AB is clearly
a poor candidate for dynamical mass measurements.

\section{Discussion}

\subsection{Is {\namesh}AB a True Wide Binary?}

The projected separation of {\namesh}AB is nearly twice that of most
late-type M and L dwarf binaries identified to date, and is comparable
to the ``wide''
binaries 2MASS~J12073346-3932539AB \citep[hereafter 2MASS~1207-3932AB]{cha04,cha05},
CFHT-Pl-18AB \citep{mar98,mar00b} and
LP~724-37AB \citep{pha05}.  However, the
separations of these systems are not necessarily inconsistent with survey results.
\citet{me03b} found that the maximum projected separations of
low mass binaries scaled with the total mass as
\begin{equation}
\rho_{max} = 1400{\times}{\rm M}_{tot}^2~{\rm AU} \label{eqn1}
\end{equation}
(see also Close et al.\ 2003).
For {\namesh}AB, our mass estimates imply $\rho_{max}$ = 27--40~AU, depending on the age of the
system.  Hence, the measured separation is close to, but not inconsistent with, the empirical limit
of Eqn~\ref{eqn1}.
Similarly, both CFHT-Pl-18AB and LP~724-37AB, with $\rho$ = 35 and 33~AU and
M$_{tot}$ $\approx$ 0.18 and 0.20~M$_{\sun}$, respectively, are well-constrained
by this relation. Hence, we do not consider any of these binaries to be
abnormally wide.

2MASS~1207$-$3932AB, however, presents a different problem.
Its membership in the TW Hydra moving association \citep{giz02} implies a young
age (10~Myr; Webb et al.\ 1999; Mamajek 2005), and correspondingly low substellar masses for its late-type M
and L dwarf components.
Current estimates of M$_{tot}$ for this system range over 0.025--0.030~M$_{\sun}$ \citep{cha04,mam05},
implying $\rho_{max} \approx 1$~AU, far less than its observed
separation of $\sim$40--55~AU.  Similarly, the 2~Myr Cha~I binary 2MASS~1101$-$7732AB,
with an estimated system mass of 0.075~M$_{\sun}$, is about 30 times wider than
the 8~AU limit implied by Eqn.~\ref{eqn1}; and the 2~Myr Taurus low-mass binary GG~TauBab
is roughly 6 times wider than the limit for its estimated system mass (0.15--0.16~M$_{\sun}$;
White et al.\ 1999; Mohanty et al.\ 2003).
All three of these systems are quite young, and \citet{mug05} have argued
(in the case of 2MASS~1207$-$3932AB) that such weakly bound systems
may not have had time to undergo the gravitational encounters that would eventually tear them apart.
On the other hand, all three are found in relative low density associations,
so such encounters may be rare.
Recent simulations by \citet{bat05} have
shown that wide low mass systems could form when two nearby but unrelated objects are ejected
from a star forming region in the same direction; however, this mechanism
is not predicted to occur in low density environments like Taurus Auriga, TW Hydra and Cha~I.

Perhaps the most interesting low mass wide binary identified thus far is DENIS~0551$-$4434AB
which, as a field source, is unlikely to have an age less than 10~Myr.
While common proper motion has not yet been established for this pair, their
angular proximity and similar spectrophotometric distances have been cited
as strong evidence
for physical association.  With a
projected separation of 220~AU, similar to GG~Tau~Bab and 2MASS~1101$-$7732AB,
this system far exceeds the separation limit
set down by Eqn.~\ref{eqn1}.
\citet{bil05} have pointed out that this system could be a higher multiple; i.e.,
either A or B or both could harbor unresolved close companions.  However,
an equal-mass quadrupole or higher is required to fit Eqn.~\ref{eqn1},
and this would imply a larger spectrophotometric distance (already estimated at 100 pc)
and hence wider projected separation.
The possibility of one or both components harboring a cool white dwarf companion
(c.f., the triple system LHS~4039, LHS~4040 and APMPM~J2354-3316C; Scholz et al.\ 2004) is
constrained by the lack of significant blue optical flux, such that this system would have
to be extremely old.  Neither of these explanations are particularly appealing.

In short, wide ($a > 100$~AU) low mass and/or brown dwarf binaries appear to be
a reality,
but as yet these systems are very young ($\lesssim$10~Myr)
and/or exceedingly rare.  Furthermore, all were identified
serendipitously, so it remains unclear as to whether they are
merely flukes of circumstance or that the tight separation limit
deduced from high resolution imaging surveys (such as Eqn.~\ref{eqn1}) is not
representative of the brown dwarf population as a whole.
Either situation has important implications on the formation
mechanism of low mass stars and brown dwarfs.  Therefore, a targeted deep imaging survey,
preferably employing proper motion corroboration, would be valuable.

\subsection{A Probe of the M dwarf/L dwarf Transition}

A robust empirical characterization of the transition between M and L dwarfs
has important reprocussions on our understanding of the atmospheres of these
objects, in which condensate dust becomes an increasingly (but possibly
variable) opacity source \citep{tsu96a,tsu96b,jon97,bur99,all01,ack01,coo03}.
By definition, the beginning of the L dwarf class
is marked by waning TiO and VO bands that characterize
M dwarfs \citep{kir99}.  These gaseous species are lost to the formation of condensates, including
titanates and mineral composites \citep{lod02}, which collectively radiate
as a warm blackbody in the near infrared.
Other condensate constituents such as Al and Ca are also seen
to disappear at the M dwarf/L dwarf boundary \citep{mcl03}.
These trends have been identified amongst samples of field dwarfs,
and as such may be blurred by the range of ages, masses and compositions
that comprise such samples.
Characterizing these trends independent
of metallicity or age effects could provide a more accurate picture of
condensate formation and other atmospheric
processes.  Under the
assumption of coevality, well-resolved M dwarf/L dwarf binaries such as {\namesh}AB enable
such studies.

In addition, chromospheric activity, as traced through H$\alpha$ emission,
is seen to decrease dramatically among the late-type M and early-type L dwarfs,
both in terms of frequency and emission strength
\citep{giz00,wes04}.  Theoretical considerations suggest that this trend
may be the result of an increasingly neutral, cool upper atmosphere, and the
subsequent impedance in transferring plasma to a hot chromosphere \citep{gel02,moh02}.
However, flaring activity does not disappear in the early L dwarfs (e.g., Hall 2002a,b),
and H$\alpha$ emission is seen even in cooler T-type brown dwarfs \citep{me00b,me03d}.
\citet{giz00} point out that older, more massive L dwarfs are more likely to exhibit
H$\alpha$ emission, suggesting a mass dependency.  Again, as prior empirical studies have
been conducted with heterogenous field samples, optical spectroscopy and monitoring
observations of resolved M dwarf/L dwarf pairs may provide a more precise
assessment of quiescent and flaring activity trends across this spectral transition.

\section{Summary}

We have resolved the field dwarf {\namesh}AB into an M9 + L0 binary
separated by 1$\farcs$09$\pm$0$\farcs$06 (38$\pm$3~AU).  Resolved
spectroscopy has enabled accurate classification of the components,
while physical association is deduced from the angular proximity, similar
spectrophotometric distances and common proper motion of the two
components.  We find that while this system,
CFHT-Pl-18AB and LP~724-37AB may be wider than
most M and L dwarf binaries discovered to date, all three remain well constrained
by empirical separation limit trends deduced from high resolution imaging surveys.
This is not the case, however, for the young wide binaries GG~Tau~Bab, 2MASS~1207$-$3932AB
and 2MASS~1101$-$7732AB, and the recently identified field pair DENIS~0551$-$4434AB.
These four systems may prove to be exceptions to
empirical separation trends, or an indication that
these trends do not adequately represent the outcome of low mass star and
brown dwarf formation.  A dedicated survey is required to address these possibilities.
Well-resolved M dwarf/L dwarf binaries such as {\namesh}AB are valuable for
studies of dust formation and
activity trends at these spectral types, and characterization of the individual
components of these systems should be done
to probe these processes in detail.

\acknowledgments

The authors would like to thank telescope operator Paul Sears and
instrument specialist John Rayner for their assistance during the
observations, and our referee Suzanne Hawley for her rapid and thorough review
of the original manuscript.  We also thank Kelle Cruz for providing SpeX spectral
data of 2MASS~0345+2540, and Michael Cushing for consultation on the
spectral data reduction.  AJB acknowledges useful discussions
with Peter Allen, Laird Close, Kelle Cruz, I.\ Neill Reid and Nick Siegler
during the preparation of the manuscript. This publication makes
use of data from the Two Micron All Sky Survey, which is a joint
project of the University of Massachusetts and the Infrared
Processing and Analysis Center, and funded by the National
Aeronautics and Space Administration and the National Science
Foundation. 2MASS data were obtained from the NASA/IPAC Infrared
Science Archive, which is operated by the Jet Propulsion
Laboratory, California Institute of Technology, under contract
with the National Aeronautics and Space Administration.

Facilities: \facility{IRTF(SpeX)}

\clearpage

\clearpage

\clearpage

\clearpage

\begin{deluxetable}{lll}
\tabletypesize{\footnotesize}
\tablecaption{Properties of {\name}AB. \label{tab1}}
\tablewidth{0pt}
\tablehead{
\colhead{Parameter} &
\colhead{Value} &
\colhead{Ref} \\
}
\startdata
$\alpha$\tablenotemark{a} & 22$^h$00$^m$02$\fs$01 & 1 \\
$\delta$\tablenotemark{a} & $-$30$\degr$38$\arcmin$32$\farcs$7 & 1 \\
$\mu$ & 0$\farcs$22$\pm$0$\farcs$05 yr$^{-1}$ & 2 \\
$\theta$ & 107$\pm$4$\degr$ & 2 \\
d  & 35$\pm$2 pc & 3 \\
$\rho$ & 1$\farcs$094$\pm$0$\farcs$003$\pm$0$\farcs$060\tablenotemark{b} & 3 \\
 & 38$\pm$3 AU & 3 \\
$\phi$ & 176$\fdg$7$\pm$0$\fdg$3$\pm$2$\fdg$0\tablenotemark{b} & 3 \\
$J$\tablenotemark{c} & 13.44$\pm$0.03 mag & 1 \\
$H$\tablenotemark{c} & 12.66$\pm$0.03 mag & 1 \\
$K_s$\tablenotemark{c} & 12.20$\pm$0.03 mag & 1 \\
${\Delta}J$ & 0.328$\pm$0.018$\pm$0.100\tablenotemark{b} mag &  3 \\
${\Delta}H$ & 0.290$\pm$0.011$\pm$0.090\tablenotemark{b} mag &  3 \\
${\Delta}K$ & 0.252$\pm$0.015$\pm$0.080\tablenotemark{b} mag &  3 \\
M$_{total}$\tablenotemark{d} & 0.14--0.17~M$_{\sun}$ &  3,4 \\
$q$\tablenotemark{d} & 0.96--0.98 &  3,4 \\
Period\tablenotemark{d} & $\sim$750--1000 yr & 3,4 \\
\enddata
\tablenotetext{a}{Equinox J2000 coordinates at epoch 1998.61 from 2MASS.}
\tablenotetext{b}{Estimates of systematic uncertainty based on PSF fitting simulations; see $\S$~2.1.}
\tablenotetext{c}{2MASS photometry of combined (unresolved) system.}
\tablenotetext{d}{Assuming an age of 0.5--5~Gyr.}
\tablerefs{(1) 2MASS \citep{cut03}; (2) SuperCosmos Sky Survey \citep{ham01a,ham01b,ham01c}; (3) This paper;
(4) \citet{bur97}.}
\end{deluxetable}

\begin{deluxetable}{lcccc}
\tabletypesize{\small}
\tablecaption{Spectral Classification Indices \label{tab2}}
\tablewidth{0pt}
\tablehead{
 & \multicolumn{2}{c}{{\namesh}A} & \multicolumn{2}{c}{{\namesh}B} \\
\cline{2-3} \cline{4-5}
\colhead{Index} &
\colhead{Value} &
\colhead{SpT} &
\colhead{Value} &
\colhead{SpT} \\
}
\startdata
\multicolumn{5}{c}{\citet{rei01}} \\
\hline
{\water}-A & 0.789  &  M8 & 0.741  & M9.5 \\
{\water}-B & 0.869  &  M9 & 0.831 & L0 \\
$K1$\tablenotemark{a} & 0.088  &  M9 & 0.125  & L0 \\
\hline
\multicolumn{5}{c}{\citet{tes01}} \\
\hline
s{\water}$^{J}$ &  0.040 &  L0.5 & 0.049 & L0.5 \\
s{\water}$^{H1}$ & 0.152  &  M9.5 & 0.203 & L0 \\
s{\water}$^{H2}$ & 0.370  &  L0.5 & 0.389 & L1 \\
s{\water}$^{K}$ &  0.074 &  M7.5 & 0.126 & M9 \\
\enddata
\tablenotetext{a}{Index defined in \citet{tok99}.}
\end{deluxetable}

\begin{deluxetable}{llccccccc}
\tabletypesize{\footnotesize}
\tablecaption{Properties of the {\name}AB Components \label{tab3}}
\tablewidth{0pt}
\tablehead{
 & & & \multicolumn{3}{c}{2MASS} & \multicolumn{3}{c}{Estimated Mass\tablenotemark{b}} \\
\cline{4-6} \cline{7-9}
\colhead{Component} &
\colhead{SpT} &
\colhead{\teff\tablenotemark{a}} &
\colhead{$J$} &
\colhead{$H$} &
\colhead{$K_s$} &
\colhead{0.5 Gyr}  &
\colhead{1 Gyr}  &
\colhead{5 Gyr} \\
\colhead{} &
\colhead{} &
\colhead{(K)} &
\colhead{} &
\colhead{} &
\colhead{} &
\colhead{(M$_{\sun}$)} &
\colhead{(M$_{\sun}$)} &
\colhead{(M$_{\sun}$)} \\
}
\startdata
{\namesh}A & M9  & 2400 & 14.05$\pm$0.10 & 13.28$\pm$0.10 & 12.83$\pm$0.10 & 0.072 & 0.079 & 0.085 \\
{\namesh}B & L0  & 2300 & 14.36$\pm$0.10 & 13.57$\pm$0.10 & 13.09$\pm$0.10 & 0.069 & 0.077 & 0.083 \\
\enddata
\tablenotetext{a}{{\teff} estimated from the {\teff}/spectral type relation of \citet{gol04}.}
\tablenotetext{b}{Based on the solar metallicity models of \citet{bur97}.}
\end{deluxetable}

\end{document}